\documentclass{article}
\usepackage[preprint]{neurips_2019}

\usepackage[utf8]{inputenc} % allow utf-8 input
\usepackage[T1]{fontenc}    % use 8-bit T1 fonts
\usepackage{hyperref}       % hyperlinks
\usepackage{url}            % simple URL typesetting
\usepackage{booktabs}       % professional-quality tables
\usepackage{amsmath}
\usepackage{amsfonts}       % blackboard math symbols
\usepackage{nicefrac}       % compact symbols for 1/2, etc.
\usepackage{microtype}      % microtypography

\usepackage{graphicx}
\usepackage[export]{adjustbox}
\usepackage[caption=false]{subfig}
\usepackage{algorithm}
\usepackage[noend]{algpseudocode}
\usepackage{amsthm}
\usepackage{rotating}
\usepackage[dvipsnames]{xcolor}

\usepackage{natbib}

\newcommand{\D}{\displaystyle}

\newcommand{\f}{\frac}

\newcommand{\bc}{\begin{center}}
\newcommand{\ec}{\end{center}}

\title{Cryptocurrency Price Prediction and Trading Strategies Using Support Vector Machines}

\author{%
  David Zhao \\
  Department of Statistics and Data Science \\
  Carnegie Mellon University \\
  Pittsburgh, PA 15213 \\
  \texttt{davidzhao@cmu.edu} \\
   \And
  Alessandro Rinaldo \\
  Department of Statistics and Data Science \\
  Carnegie Mellon University \\
  Pittsburgh, PA 15213 \\
  \texttt{arinaldo@cmu.edu} \\
  \AND
  Christopher Brookins \\
  Valiendero Digital Assets, LP \\
  Pittsburgh, PA 15213 \\
  \texttt{chris@valiendero.io} \\
}

\begin{document}

\maketitle

\begin{abstract}
Few assets in financial history have been as notoriously volatile as cryptocurrencies. While the long term outlook for this asset class remains unclear, we are successful in making short term price predictions for several major crypto assets. Using historical data from July 2015 to November 2019, we develop a large number of technical indicators to capture patterns in the cryptocurrency market. We then test various classification methods to forecast short-term future price movements based on these indicators. On both PPV and NPV metrics, our classifiers do well in identifying up and down market moves over the next 1 hour. Beyond evaluating classification accuracy, we also develop a strategy for translating 1-hour-ahead class predictions into trading decisions, along with a backtester that simulates trading in a realistic environment. We find that support vector machines yield the most profitable trading strategies, which outperform the market on average for Bitcoin, Ethereum and Litecoin over the past 22 months, since January 2018.
\end{abstract}

\section{Introduction}
\label{introduction}
The price of Bitcoin has swung wildly over the past several years, drawing interest from both economists trying to fundamentally understand the phenomenon (\cite{EconomicsBitcoinPrice} \cite{Athey}) and investors trying to simply predict price swings in order to earn profits (\cite{StackedNNBitcoin} \cite{DeepRLBitcoin} \cite{CryptoassetFactorModels}). This paper falls into the latter camp. We abstract away the economic and political contexts in which cryptocurrencies exist, focusing solely on the problem of future price prediction given past trading data. Even without considering first principles, it is a daunting task to build an explicit model that accounts for the many possible sources of variation for a poorly understood asset like Bitcoin. Such a model is likely to be overly complex and not useful for prediction. Instead, in this paper we opt for a black-box approach that simply targets prediction accuracy and backtest profit in a time-varying setting.

\subsection{What is Bitcoin?}
Bitcoin is a peer-to-peer payment system introduced as open source software in January 2009 by a computer programmer using the pseudonym Satoshi Nakamoto (\cite{Nakamoto}). Bitcoin utilizes cryptographic principles in transaction verification, privacy, and issuance. Key features of the Bitcoin system are having no central point of trust or failure (decentralization), not being confined to any local jurisdiction or fiat currency (global reach), a fixed money supply (only 21 million \$BTC will ever be issued), divisibility and fungibility, and lower transaction costs than traditional alternatives. 
\\\\
Bitcoin has become a worldwide phenomenon and has experienced incredible growth, leaping from a market capitalization of essentially zero to over \$114 billion (at time of writing) in less than 10 years. Coupled with this parabolic growth has been Bitcoin's detachment from macroeconomic features that have been widely tracked for traditional financial asset pricing. Table \ref{table:corr} shows that Bitcoin has very little correlation with major traditional financial assets (\cite{Corrs}).
\begin{table}[h!]
	\begin{center}
		\caption{\textit{Unconditional pairwise Pearson correlations of Bitcoin with major financial assets over the period Jul 2, 2011 to Dec 31, 2017. See \cite{Corrs} for source.}}
		\label{table:corr}
		\begin{tabular}{l|c}
			\textbf{Asset} & \textbf{Correlation with Bitcoin}\\
			\hline
			Gold & 0.0459 \\
			Silver & 0.0071 \\
			WTI (Crude Oil) & 0.0158 \\
			S\&P 500 & 0.0491 \\
			MSCI World & 0.0457 \\
			MSCI EM50 & 0.0042 \\
		\end{tabular}
	\end{center}
\end{table}
\\\\
The nascency of this market, its noise and extreme volatility, decorrelation with existing assets, rapidly shifting sentiment, lack of established fundamental economic valuation, and even suspected price manipulation (\cite{PriceManipulation}) make Bitcoin one of the hardest prediction problems to solve. Investors with diverse strategies can be found across the spectrum. Some are purely fundamental investors in the venture capital style, focusing on the technological implications of the blockchain. (The blockchain is an immutable public ledger that records all transactions for a given cryptocurrency. We describe blockchain data in more detail in Section \ref{onchaindata}.) Some are applying high frequency trading strategies that are widely utilized in traditional markets. Some are mining sentiment data from news and social network platforms to predict market cycles.
\\\\
Our approach is simple yet effective at beating the market over an almost 2-year period for Bitcoin, as well as ``altcoins'' Ethereum (\cite{Ethereum}) and Litecoin (\cite{Litecoin}), which are among the most widely used cryptocurrencies with the highest capitalizations, next to Bitcoin. In Section \ref{description}, we describe the datasets we work with and features we generate. In Section \ref{methodology}, we discuss our methodology for generating price predictions and translating them into trading decisions. In Section \ref{results}, we present our results on classification accuracy and backtested profit of our strategies.

\section{Data and Features}
\label{description}
We scrape publicly available historical price and volume data from Coinbase (\url{https://coinbase.com/}), which is the world's largest Bitcoin (BTC) broker, as well as a platform for trading numerous other cryptocurrencies such as Ethereum (ETH) and Litecoin (LTC). We also scrape on-chain data, which describes the blockchain i.e. the underlying cryptographic network for crypto assets, via Blocktap (\url{https://www.blocktap.io/}). From this data, we develop a large number of technical indicators that serve as input features for our classification algorithms.

\subsection{Coinbase data}
\label{coinbasedata}
\begin{figure}
	\centering
	\includegraphics[scale=0.7]{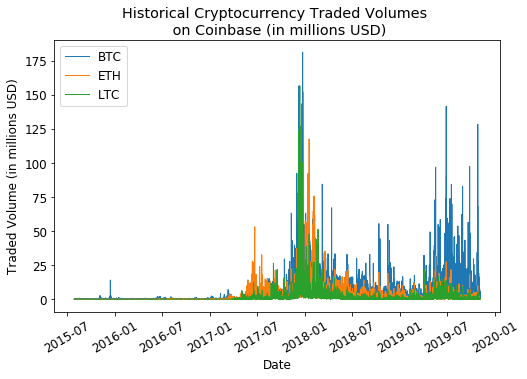}
	\caption{\textit{Historical volumes for BTC, ETH, and LTC from July 2015 to the present are highly volatile, as well as highly correlated with each other and with historical prices.}}
	\label{fig:historical_volumes}
\end{figure}

Coinbase provides \texttt{open}, \texttt{high}, \texttt{low}, \texttt{close}, and \texttt{volume} data (OHLCV) at 1 hour intervals for the currency pairs BTCUSD, ETHUSD, and LTCUSD. \texttt{open} and \texttt{close} refer to the asset price at the start and end of the interval, respectively. \texttt{high} is the highest price the asset attains within the interval, while \texttt{low} is the lowest price within the interval. \texttt{volume} is the total volume, in units of cryptocurrency, of trades made on the Coinbase exchange within the interval. Historical data for BTCUSD is available at 1 minute intervals since July 2015. ETHUSD is available at 1 minute intervals since Aug 2016, and LTCUSD is available at 1 minute intervals since Sep 2016.
\begin{sidewaysfigure}[]
	\centering
	\includegraphics[scale=0.8]{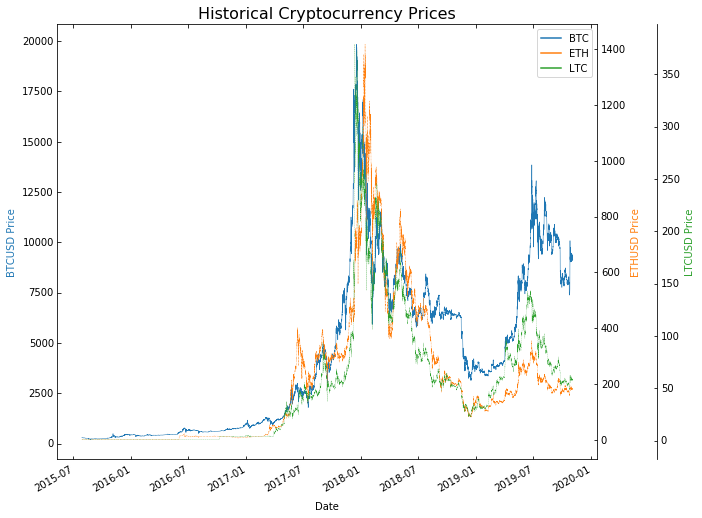}
	\caption{\textit{Historical prices for BTC, ETH, and LTC from July 2015 to the present are highly volatile, as well as highly correlated with each other.}}
	\label{fig:historical_prices}
\end{sidewaysfigure}
\\\\
Figures \ref{fig:historical_prices} and \ref{fig:historical_volumes} show that the prices and traded volumes of these cryptocurrencies have fluctuated dramatically over the past several years. Moreover, the cross-asset time series correlations are all high, and price and volume tend to be highly correlated as well.
\\\\
We form various technical features based on historical prices and volumes. In Appendix A, we describe a few illustrative examples, to give context for the kinds of features one might derive from historical market data. The basic idea is we attempt to capture the ``direction'' prices have been moving in the recent past. We implicitly assume that recent price patterns do in fact influence future prices, and that our features capture relevant properties of recent price evolutions.

\subsection{On-chain data}
\label{onchaindata}
Blocktap is a data platform that provides on-chain data, which refers to all data for a digital asset that is natively stored on that asset's public blockchain. Blockchain data differs by asset, because each asset has its own unique use case or terminology. Thus, blockchain metrics may differ in title and significance for price prediction, asset to asset. Some examples of on-chain data terminology include miners revenue, mining rate, gas, active addresses, native coin emission rate, transaction count, transaction volume, block size, and more. Furthermore, each metric represents some fundamental economic component of the asset's network value, i.e. supply or demand. For example, active addresses generally denote a demand side feature while native coin emission rate is supply side.
\\\\
The largest, free data repositories of such information can be found at \url{coinmetrics.io} and \url{blocktap.io}, whereas real-time transaction tracking and metric graphics can be found at \url{blockchain.com} (Bitcoin blockchain) and \url{etherscan.io} (Ethereum blockchain). These information sources provide a host of metrics ranging from user adoption rate to investor purchases to recent transactions. This type of open source and real-time insight into a financial economy and its players' movements is the first of its kind. Furthermore, it is hypothesized that certain on-chain data metrics, such as stock-to-flow (supply of BTC vs. how much new BTC coming out), influence the underlying price of digital assets like Bitcoin. As the asset class matures, we expect greater utilization of on-chain data into predictive models as we have explored in this paper.

\section{Methodology}
\label{methodology}
First, we define price prediction as a classification problem and propose classifiers to solve it, using various features we derive from market and on-chain data, as described in Section \ref{description}. Beyond generating price predictions, we develop a framework for backtesting how well our strategy would have performed, in a quasi-realistic setting, over the period of available data.

\subsection{Problem Formulation}
\label{problem}
\begin{figure}[h!]
	\centering
	\subfloat{
		\includegraphics[scale=0.4, valign=t]{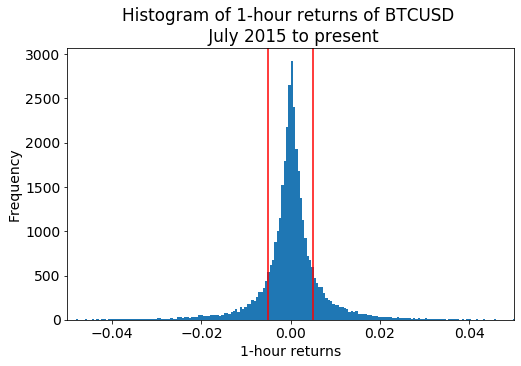}
	}
	\subfloat{
		\includegraphics[scale=0.4, valign=t]{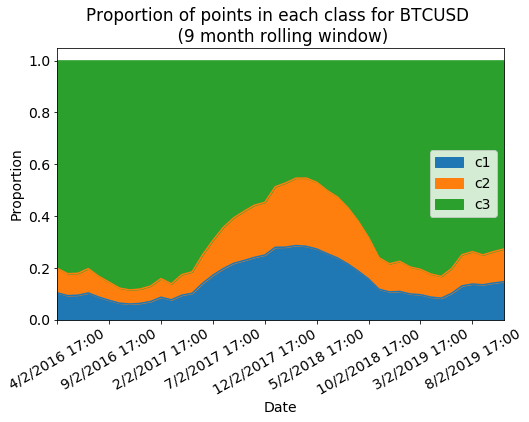}
	}
	\caption{\textit{LEFT: The distribution of 1-hour returns for BTCUSD from July 2015 to the present. The two vertical lines indicate the thresholds for class membership, at 0.005 and -0.005. RIGHT: We show the 9 month rolling proportion of points in each class, from 2016 to the present, for BTCUSD. We will use this 9 month rolling window to retrain our classifiers, as described in Section \ref{SVM}. These classes are highly imbalanced, with the majority of points falling in class $c_3$. We clearly see a higher proportion of the classes $c_1$ and $c_2$ during the volatile bubble and crash periods of 2017-2018.}}
	\label{fig:returns}
\end{figure}

\begin{figure}[h!]
	\centering
	\subfloat{
		\includegraphics[scale=0.5]{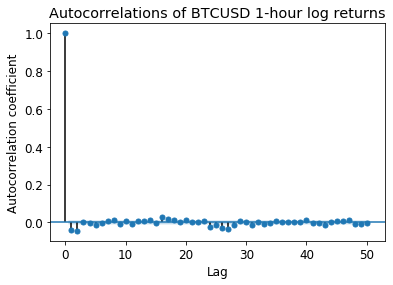}
	}
	\subfloat{
		\includegraphics[scale=0.5]{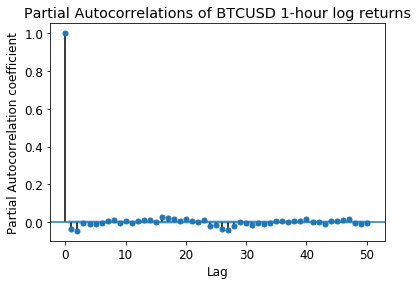}
	}
	\caption{\textit{The first 50 autocorrelations and partial autocorrelations of 1-hour log returns for BTCUSD are all close to zero. ETHUSD and LTCUSD exhibit similar properties.}}
	\label{fig:autocorr}
\end{figure}

Our response variable is the 1-hour-ahead percentage change in log price:
$$\D r_{t+1} = \f{\log P_{t+1} - \log P_t}{\log P_t}$$
Taking the logarithm removes the unit root present in currency prices (\cite{UnitRoot}). (We see in Figure \ref{fig:autocorr} that the series of log returns is whitened.) First, we discretize the response, turning this regression problem into a classification problem. We define 3 classes:
\bc
	$c_1$: ``up'', $r_t \geq 0.5\%$; \ \ \  
	$c_2$: ``down'', meaning $r_t \leq -0.5\%$; \ \ \ 
	$c_3$: ``same'', otherwise
\ec
Figure \ref{fig:returns} shows the distribution of 1-hour log-returns. Two vertical lines at -0.005 and 0.005 indicate the thresholds for class membership. We believe this threshold of 50 basis points is a reasonable, conservative choice, given the cost of transactions on Coinbase (see \ref{fees}). It would of course not be practically useful to predict price moves too small to cover the transaction costs of trading.
\\\\
Since our goal is prediction, not inference, we incorporate historical information directly into predictive features, instead of explicitly fitting a time series model. Moreover, Figure \ref{fig:autocorr} shows the autocorrelations and partial autocorrelations of 1-hour log returns are all close to zero.

\subsection{Classification Methods}
\label{SVM}
Explicitly specifying a high-dimensional parametric model is not only prohibitively complex, but prediction intervals would also tend to be wide and unusable, given the inherent noisiness of the data generating process. Further, any such parametric model would require many assumptions (e.g. stationarity) that we have no reason to believe are true. All things considered, a black-box approach seems appropriate. We try several multiclass classification methods, including support vector machines (SVM), random forest, and XGBoost. We train these classifiers to predict $r_{t+1}$ given all the relevant features from Coinbase and Blocktap at time $t$. Since our dataset is sometimes very imbalanced (vast majority of returns in a 1-hour period fall into a single class, usually $c_3$), we weight classes inversely by their frequency in the dataset. We find that SVM with a squared hinge loss is our best performing classifier. See Appendix B for a detailed explanation of SVM.
\\\\
We must also decide how much historical data to use to train our classifiers. Including too much history runs the risk of training on irrelevant information, as the present regime may differ completely from the past. Including too little history runs the risk of underfitting the classifiers due to inadequate data. We decide to use a rolling window of 9 months, which we believe is a sensible balance between training on enough data while avoiding unhelpful data from the distant past. We also find empirically that using a 9 month window performs well. Every month, we retrain the classifier on the most recent 9 months of data. Taking this classifier as fixed over the course of the next month, every hour we use the classifier to generate 1 hour ahead predictions based on the current data that hour. At the end of the next month, we retrain the classifier, then generate new hour-by-hour predictions for the following month. Empirically, we did not see any benefit to constantly refitting the classifiers with the latest hour of information$-$updating the models monthly was sufficient.

\subsection{Gamma thresholding}
As described in Section \ref{SVM}, every hour, SVM generates a class prediction for the return over the next hour. These predictions often are not ``confident'' or significant enough to warrant any action. Put another way, we would rather make fewer trading decisions that are very accurate, in lieu of many trades that on average are only slightly better than chance. The former scenario is more profitable, especially after considering the trading costs a small, non-institutional player would incur (see Section \ref{fees}). To address this, we essentially use a form of classification with rejection (\cite{ClassReject}), saying ``we don't know'' when the classifier does not seem ``confident'' enough in its prediction. 
\\\\
Formally, we define a tuning parameter $\gamma$, which serves as a threshold for making trading decisions. In the case of SVM, this tuning parameter is a threshold for the signed distance to the margin (See Appendix B.2). In other words, we will only consider potentially making a buy trade (i.e. consider a new point to belong to class $c_1$) if SVM predicts $c_1$ \textit{and} the estimated distance to the margin exceeds $\gamma$. Random forest and XGBoost are perhaps more intuitive examples$-$here, $\gamma$ will serve as a threshold for the predicted class probability from the regression functions. For three classes, a naive Bayes' rule would set $\gamma=1/3$ and suggest buy trades whenever the predicted probability weight on $c_1$ exceeds 1/3. If we choose a higher $\gamma > 1/3$, we effectively make our trading strategy more conservative, setting a higher bar to meet before trading actions are taken. Empirically, we find that using a $\gamma$ threshold for predicting classes $c_1$ and $c_2$ improves NPV and PPV (see Figure \ref{fig:PPV_NPV}), and in turn profitability (see Section \ref{bitcoin}).

\subsection{Backtesting}
\label{backtesting}
We translate class predictions into trading decisions, and then we develop a backtester to simulate in a quasi-realistic setting how well our trading strategy would have done historically.

\subsubsection{Trading Schedule}
\label{schedule}
To simplify things, we assume we always hold either 100\% cash or 100\% cryptocurrency (only allowed to take long positions). If we are holding all cash, every hour we evaluate whether the predicted class is $c_1$ and the signed distance to the margin exceeds $\gamma$. If both conditions are met, we convert all our cash into cryptocurrency, buying at the current market price (we enter into a long position). If we are holding all cryptocurrency, every hour we evaluate whether the predicted class is $c_2$ and the signed distance to the margin exceeds $\gamma$. If both conditions are met, we convert all our cryptocurrency into cash, selling at the current market price (we exit the long position).
\\\\
Algorithm \ref{tradingalgo} describes our trading strategy. Recall that we retrain the SVM classifier every month, using a rolling 9 month window, and we use it to make hourly predictions over the next month, denoted $\hat{y}_1,\ldots,\hat{y}_T$ in the algorithm below. We can of course refine this strategy by allowing shorting, as well as position sizing based on holdings and prediction confidence. But even the crude ``all-in'' strategy described here can outperform the market, as we will see in Section \ref{bitcoin}.
\begin{algorithm}
	\caption{\textit{Long-only trading strategy}}
	\label{tradingalgo}
	\begin{algorithmic}[1]
		
		\State \textbf{Input:} SVM class predictions $\hat{y}_1,\ldots,\hat{y}_T$, SVM signed distances to margin $d_1,\ldots,d_T$, bitcoin prices $P_1,\ldots,P_T$, threshold $\gamma \geq 0$, take-profit $s_G > 0$, stop-loss $s_L > 0$, cash $C$
		\State set bitcoins $B=0$, portfolio value $V = C$
		\For{$t=1,\ldots,T$}		\Comment{loop over each 1-hour interval}
		\If{$C > 0$}	\Comment{currently holding all cash}
		\If{$\hat{y}_t = c_1$ AND $d_t \geq \gamma$}
		\State $\D B=\f{C}{P_t}$,\ \ \ $V=C$,\ \ \ $C=0$	\Comment{convert all cash into bitcoin}
		\EndIf
		\Else 	\Comment{currently holding all bitcoin}
		\If{$\hat{y}_t = c_2$ AND $d_t \geq \gamma$}
		\State $\D C=B \cdot P_t$,\ \ \ $B=0$,\ \ \ $V=C$	\Comment{convert all bitcoin into cash}		
		\EndIf
		\\
		\If{$\D \f{B \cdot P_t - V}{V} \geq s_G$\ \ \  OR\ \ \ $\D \f{B \cdot P_t - V}{V} \leq s_L$}	\Comment{take-profit or stop-loss breached}
		\State $\D C=B \cdot P_t$,\ \ \ $B=0$,\ \ \ $V=C$	\Comment{convert all bitcoin into cash}		
		\EndIf
		\EndIf
		\EndFor \\
		\Return
	\end{algorithmic}
\end{algorithm}

\subsubsection{Fees, Limits, and Risk}
\label{fees}
Our strategy is slow-moving enough and the market is active enough that we do not need market orders$-$limit orders at the best available price should execute within minutes, incurring a fee of 0.00\% to 0.15\% of the trade value to Coinbase Prime (\url{https://prime.coinbase.com/}). To be conservative, we assume trading costs of 0.25\% for each trade. This is because in addition to fees explicitly paid to the exchange, trading also incurs invisible fees due to market impact (\cite{MarketImpact}). The larger and more quickly one trades, the more expensive it is because one pushes the price further away.
\\\\
Our backtester does not allow leverage, so the position size is limited to 100\% of the portfolio value. We also include take-profit and stop-loss features to manage risk and modulate the aggressiveness of our strategy. These define the largest positional gain and loss, respectively, that one is willing to see before closing the position, either to lock in gains or prevent further losses.

\section{Experimental Results}
\label{results}
In both classification accuracy and backtest performance, SVM was our best model. Indeed, a trading strategy based on SVM beats the market on average from January 2018 to present. 

\subsection{Classification accuracy}
\label{accuracy}
\begin{figure}[h!]
	\centering
	\subfloat{
		\includegraphics[scale=0.8]{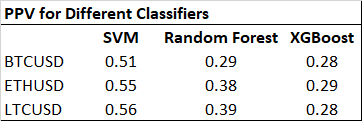}
	}\ \ \ \ \ \ 
	\subfloat{
		\includegraphics[scale=0.8]{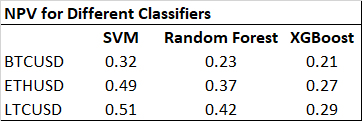}
	}
	\caption{\textit{PPV and NPV values for our classifiers, averaged from Jan 2018 to present.}}
	\label{fig:PPV_NPV}
\end{figure}

Accuracy in key decision moments matters more than overall classification accuracy, especially as our strategies only decide to trade a small fraction of the time. The overall accuracy, taken over every hour of every day, is more indicative of ``missed opportunities'' and less relevant for our trading style. (See Appendix C for more detailed results.) We focus instead on PPV and NPV, which reflect how often our model is correct among the times that it chooses to actively buy or sell.
\\\\
The positive predictive value (PPV) metric is defined as follows:
$$ PPV = \f{\textrm{\# of true positives}}{\textrm{\# of true positives + \# of false positives}} = \f{\textrm{\# of true positives}}{\textrm{\# of predicted positives}} $$
The negative predictive value (NPV) metric is defined as follows:
$$ NPV = \f{\textrm{\# of true negatives}}{\textrm{\# of true negatives + \# of false negatives}} = \f{\textrm{\# of true negatives}}{\textrm{\# of predicted negatives}} $$
\\\\
Figure \ref{fig:PPV_NPV} shows the PPV and NPV for SVM, random forest, and XGBoost. Using the decision framework described in Section \ref{backtesting}, we set a threshold $\gamma$, which means the prediction ``confidence'' must exceed $\gamma$ for the algorithm to treat it as positive (class $c_1$) or negative (class $c_2$). SVM consistently has the highest accuracy, and it forms the basis of a viable trading strategy.

\subsection{Trading decisions}
\begin{sidewaysfigure}[]
	\centering
	\includegraphics[scale=0.6]{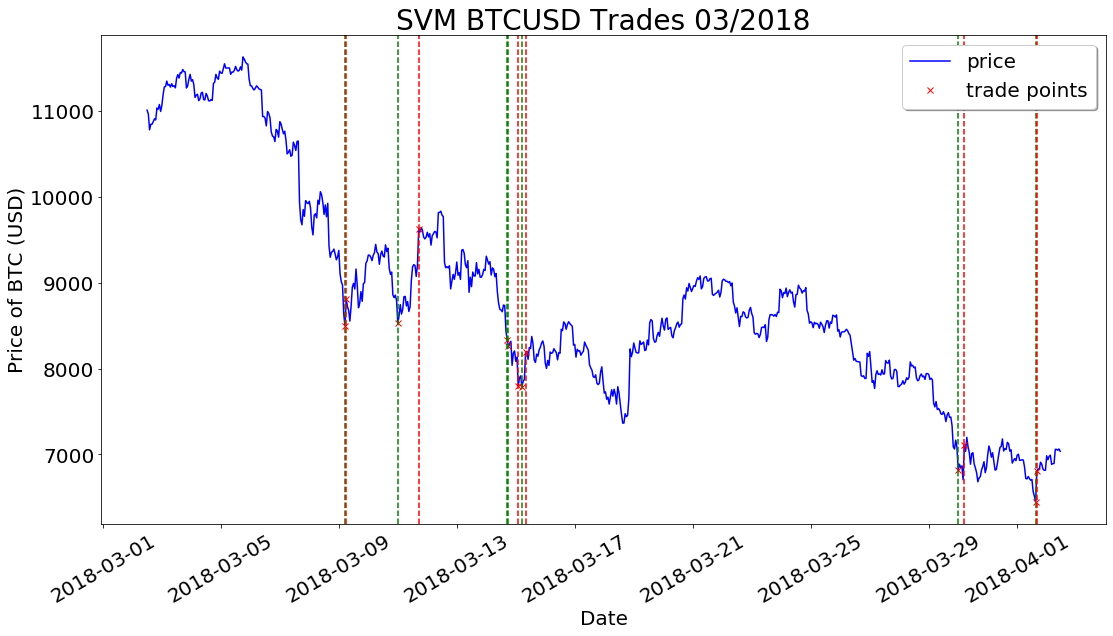}
	\caption{\textit{This figure shows that trades that the SVM model makes during Mar 2018 for BTCUSD. Red x's denote trading points, with vertical lines for visual clarity. A green line denotes a buy trade, and a red line denotes a sell trade. Visually, these trading decision points seem to correspond to sudden ``jumps'' in the price series, consistent with the momentum-based features our model uses. In a month when the price of BTC drops by 35.6\%, our algorithm, over the course of 6 pairs of buy/sell trades, returns 22.7\% after trading costs.}}
	\label{fig:backtest_example}
\end{sidewaysfigure}

Figure \ref{fig:backtest_example} shows our trading strategy in action in the BTCUSD market in March 2018. Red x's with vertical lines denotes trades, with green lines meaning buy and red lines meaning sell. Table \ref{table:trades} summarizes the 12 trades that were made. We see this was a particularly good month, where 5 out of 6 buy / sell pairs were profitable, and our algorithm made 22.7\% after trading fees while the market dropped by 35.6\%. It is also encouraging to see that visually, our trading decisions occur after sudden ``jumps'' in the price series, as expected for a strategy trading on momentum-based features.
\\
\begin{table}[h!]
	\begin{center}
		\caption{\textit{Trading decisions in BTCUSD market over month of March 2018.}}
		\label{table:trades}
		\begin{tabular}{c|c|c}
			\textbf{Type} & \textbf{Time} & \textbf{Unit Price} \\
			\hline
			\color{ForestGreen} buy & 2018-03-09 05:00:00 & 8499.90 \\
			\color{red} sell & 2018-03-09 06:00:00 & 8815.08 \\
			\color{ForestGreen} buy & 2018-03-11 00:00:00 & 8529.96 \\
			\color{red} sell & 2018-03-11 17:00:00 & 9631.78 \\
			\color{ForestGreen} buy & 2018-03-14 17:00:00 & 8335.12 \\
			\color{red} sell & 2018-03-15 02:00:00 & 7797.50 \\
			\color{ForestGreen} buy & 2018-03-15 05:00:00 & 7791.95 \\
			\color{red} sell & 2018-03-15 08:00:00 & 8194.61 \\
			\color{ForestGreen} buy & 2018-03-30 00:00:00 & 6815.01 \\
			\color{red} sell & 2018-03-30 05:00:00 & 7110.39 \\
			\color{ForestGreen} buy & 2018-04-01 15:00:00 & 6450.01 \\
			\color{red} sell & 2018-04-01 16:00:00 & 6805.01 \\
		\end{tabular}
	\end{center}
\end{table}

\subsection{Backtest results}
\label{bitcoin}
\begin{figure}[h!]
	\centering
	\subfloat{
		\includegraphics[scale=0.8, valign=t]{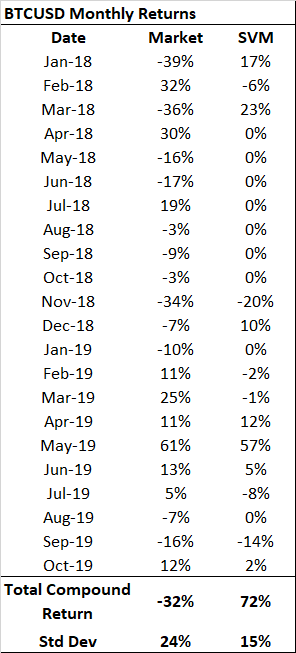}
	}
	\subfloat{
		\includegraphics[scale=0.5, valign=t]{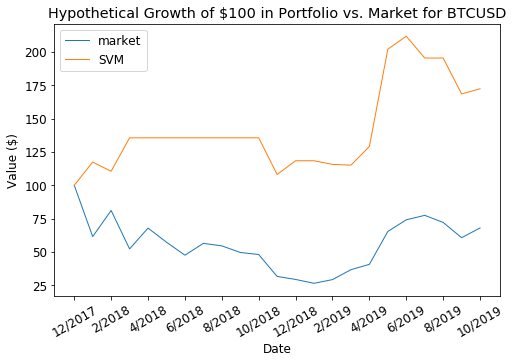}
	}
	\caption{\textit{This table shows monthly backtest returns for SVM for Bitcoin from Jan 2018 to the present, along with the total compounded return and standard deviation of returns. SVM outperforms the market somewhat consistently, while having lower volatility. The return correlation of SVM to the market is 35.5\% over this period. The plot shows these returns cumulated over the backtest period, i.e. the hypothetical growth of \$100 in our portfolio vs. the BTC market.}}
	\label{fig:BTC_backtest}
\end{figure}

\begin{figure}[h!]
	\centering
	\subfloat{
		\includegraphics[scale=0.8, valign=t]{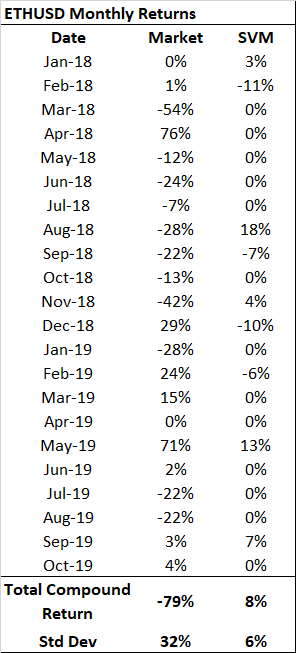}
	}\ \ \ \ \ \ \ \ \ \ 
	\subfloat{
		\includegraphics[scale=0.8, valign=t]{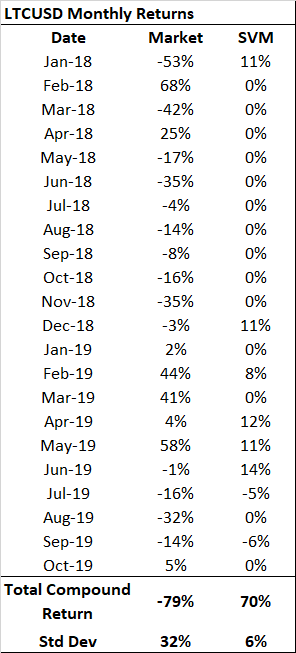}
	}
	\caption{\textit{We show monthly backtest returns for SVM for Ethereum and Litecoin from Jan 2018 to the present. For both assets, SVM outperforms the market while having lower volatility. The return correlation of SVM to the market over this period is 0.1\% for ETHUSD, and 18.5\% for LTCUSD.}}
	\label{fig:ETH_LTC_table}
\end{figure}

\begin{figure}[]
	\centering
	\includegraphics[scale=0.4]{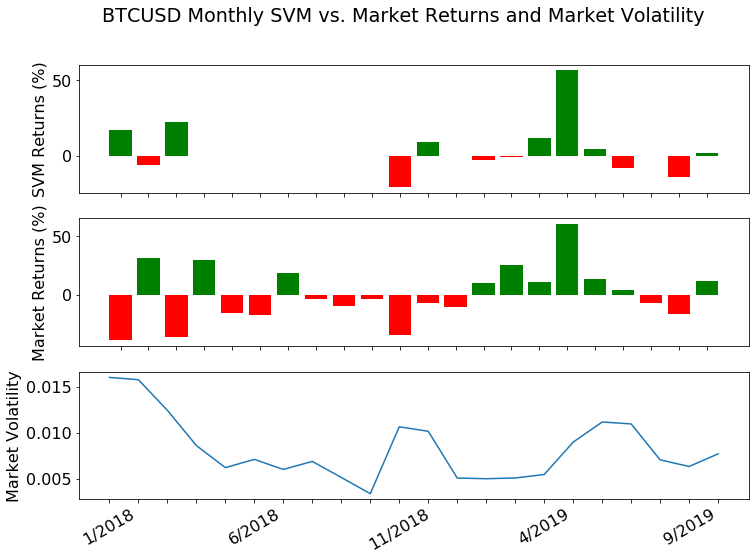}
	\caption{\textit{We compare the monthly returns for SVM and the market in BTCUSD, along with rolling market volatility. We see that our strategy is more likely to trade when market volatility is high, and often does not trade at all in low volatility periods, when it is more ``uncertain'' of any price moves. (We explore this further in Appendix D.) Our strategy is also relatively uncorrelated with the market.}}
	\label{fig:weekly_returns}
\end{figure}

Figure \ref{fig:BTC_backtest} shows monthly backtest results for BTCUSD. Each month begins on the 5th and ends on the 5th, and SVM returns are reported net of trading costs. We focus on the period Jan 2018 to the present as a separate ``regime'' from earlier years. Prior to 2018, crypto markets were in their infancy, with perhaps essentially different dynamics than the recent market, after the crypto bubble burst at the end of 2017. Our features seem work well over this recent period. Figure \ref{fig:ETH_LTC_table} shows monthly backtest results for ETHUSD and LTCUSD. Similar to Bitcoin, SVM outperforms the market somewhat consistently for Ethereum and Litecoin as well, while enjoying lower volatility.
\begin{figure}[h!]
	\centering
	\subfloat{
		\includegraphics[scale=0.4, valign=t]{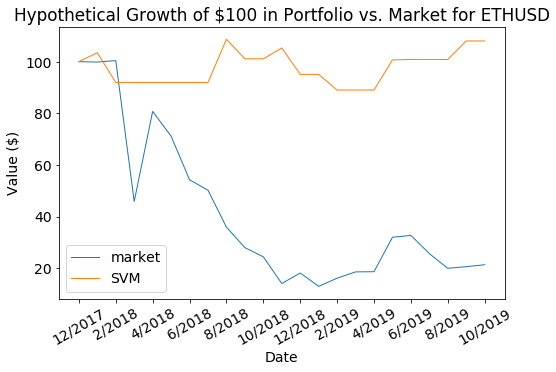}
	}
	\subfloat{
		\includegraphics[scale=0.4, valign=t]{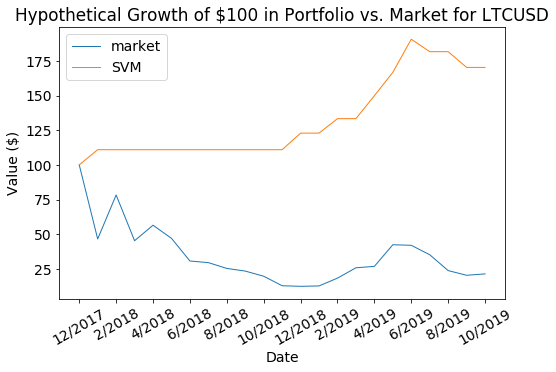}
	}
	\caption{\textit{Hypothetical growth of \$100 in our portfolio vs. the market, for ETH and LTC.}}
	\label{fig:ETH_LTC_cumulative}
\end{figure}
\\\\
These results are encouraging, as our naive ``all-in'' strategy performs well even without shorting or position sizing. Our average return over the 22 months from Jan 2018 to the present is higher than the market's, while our fluctuations and drawdowns are smaller. This is true for all three assets we tested on: BTC, ETH, and LTC. It seems that in uncertain conditions when the market declines, such as the second half of 2018, our algorithm often does not trade at all, which helps overall performance while reducing overall volatility. In all 3 crypto markets, we are able to avoid the catastrophic losses of the market, while nimbly finding opportunities to make short term profits.
\\\\
Figure \ref{fig:weekly_returns} compares the time series of monthly returns from our strategy, rolling market volatility, and market returns for BTCUSD. In all 3 crypto markets we tested, our strategy seems to trade more when volatility is higher. (We study this in more detail in Appendix D.) Furthermore, its returns are relatively uncorrelated with the market, which is a great feature that allows one to, say, hold a long cryptocurrency position while using our strategy to trade short term around that position.

\section{Discussion}
\label{discussion}
Cryptocurrency price prediction is a difficult, open-ended problem with many possible approaches. We chose a black-box classification approach over the ``medium term'', on the order of hours$-$neither ultra-short term like high frequency strategies, nor long term like venture capital investing. We gathered data from Coinbase and Blocktap, related to historical prices and volumes as well as on-chain metrics underlying the cryptographic networks. We then defined features to feed into a classification algorithm, and translated the resulting predictions into a trading strategy. While there remains much to be optimized and improved, it is heartening to see that our simple, straighforward black-box approach can deliver superior returns over an extended period of time.

\subsection{Possible extensions}
\label{extensions}
We believe there are opportunities in the cryptocurrency markets at other time horizons, both shorter and longer. For example, Coinbase provides order book information available at 1 minute intervals, which may be useful for higher frequency trading strategies. Instead of regular intervals, we could try retraining the classifier in an online fashion using change point detection (\cite{Changepoint}). We could also adjust risk limits and position sizing dynamically based on market volatility. Another idea is to train only on short periods before and after major price spikes and drops, to try to understand these key events. We could also define the classification problem differently, or choose a regression approach instead. More powerful algorithms such as ensemble methods and recurrent neural networks might improve predictions. Also, the authors initially explored conformal methods for more cautious prediction intervals (\cite{CautiousDeeplearning}), and it may prove fruitful to apply these ideas to our current framework. Outside of historical prices and volumes, sentiment data, in the form of news articles and social media posts, might reflect or influence investor sentiment and in turn influence prices. Additionally, macroeconomic factors may play an important role in cryptocurrency dynamics. Political and currency instability around the world, for example, might increase the demand for cryptoassets, which are relatively uncorrelated with the macroeconomy. Overall, we believe there is ample structure and opportunity in this relatively new, exciting asset class, and we are encouraged by the success of our simple black-box approach.

\section*{Acknowledgments}
Firstly, I am grateful to my wonderful advisor Alessandro Rinaldo for his direction, guidance and support. I am also grateful to Christopher Brookins of Valiendero Digital Assets, LP for his wisdom and abundant industry experience. It has been rewarding to collaborate with dedicated colleagues Niccolo Dalmasso, Riccardo Fogliato, and Wanshan Li.

\newpage

\bibliography{ADA_references}
\bibliographystyle{plainnat}

\newpage

\section*{Appendix A. Momentum-based features}
Below, we give a few illustrative examples of the kinds of features we might derive from historical price and volume data. In particular, we seek features that capture the general ``direction'' or ``momentum'' of the market recently. The features below by no means represent an exhaustive list of the features used in our experiments in this paper.

\subsection*{A.1. Bollinger bands}
\begin{figure}[h!]
	\centering
	\includegraphics[scale=0.4]{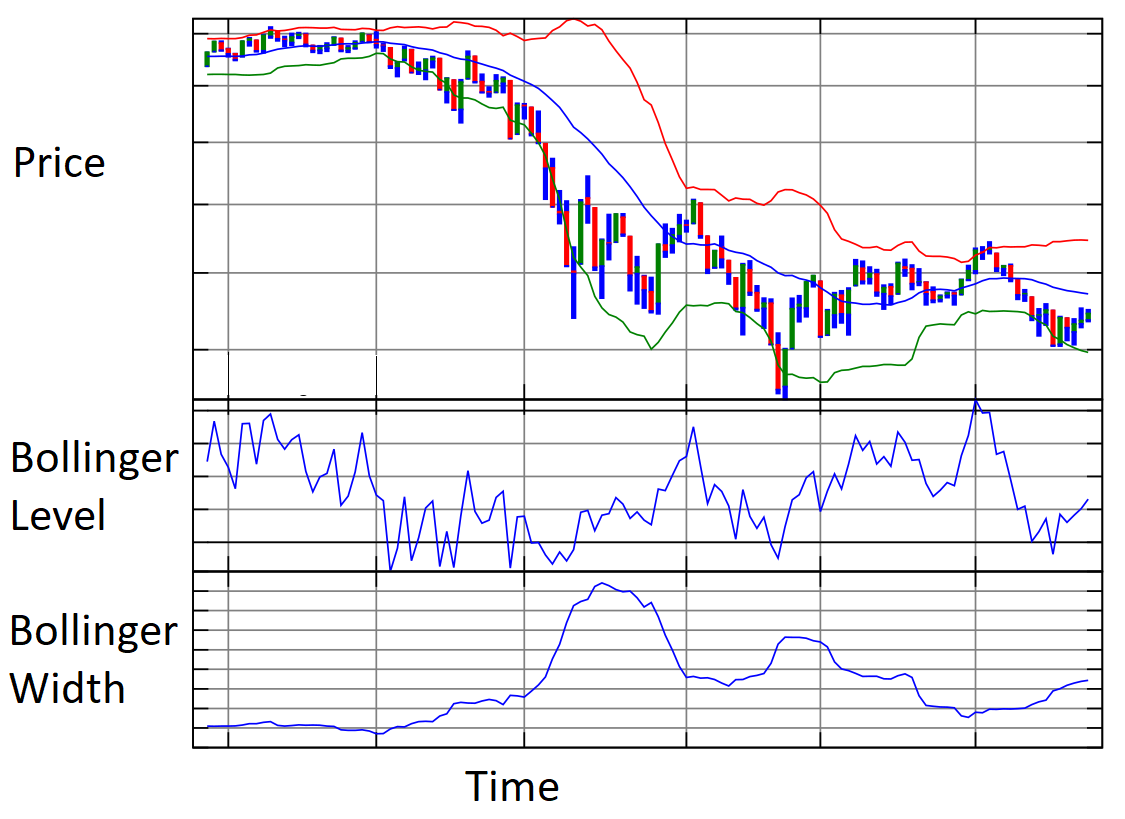}
	\caption{\textit{Here is an illustration of Bollinger bands, the Bollinger level (current price relative to upper and lower bands), and the Bollinger width (rolling volatility).}}
	\label{fig:bollinger}
\end{figure}

Bollinger bands (\cite{Bollinger}) compare the current price to rolling mean and volatility. Denote $P(i)$ as the price at time $i$, and $MA(n,t)$ as the $n$-day moving average price at time $t$. Define upper and lower bands:
$$ U(t) = MA(n,t) + 2 \sqrt{\sum_{i=t-n+1}^t \f{(P(i) - MA(n,t))^2}{n} } $$
$$ L(t) = MA(n,t) - 2 \sqrt{\sum_{i=t-n+1}^t \f{(P(i) - MA(n,t))^2}{n} } $$
Two features we derive from these Bollinger bands are \%b and bandwidth. At time $t$, we write:
$$ \%b = \f{P(t) - L(t)}{U(t) - L(t)} $$
$$ \textrm{bandwidth} = \f{U(t) - L(t)}{MA(n,t)} $$
Figure \ref{fig:bollinger} shows an example of Bollinger bands. Intuitively, we measure rolling mean and volatility in order to compare the current price level to levels in the recent past.

\subsection*{A.2. Moving average convergence-divergence}
\begin{figure}[h!]
	\centering
	\includegraphics[scale=1.0]{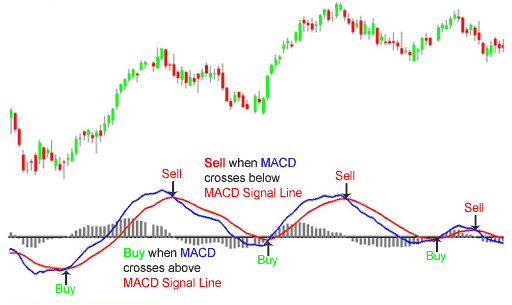}
	\caption{\textit{Here is an illustration of the various features of the moving-average convergence divergence (MACD) metric. The blue line is the MACD line, which is the fast EMA minus the slow EMA. The red line is the signal line, which is an EMA of the MACD line. The histogram line is represented by the gray histograms. This is computed as the MACD line minus the signal line.}}
	\label{fig:macd}
\end{figure}

The moving average convergence-divergence (MACD) metric (\cite{MACD}) compares two exponential moving averages (EMA) with different windows, and basically measures the level of the faster one relative to the slower one. Define the MACD line as
$$ M(t) = EMA(P,m,t) - EMA(P,n,t),\ m < n $$
where $P$ denotes price. Typical values for $m$ and $n$ are 12 and 26, respectively. Define the signal line as an exponential moving average of the MACD line:
$$ S(t) = EMA(M,l,t) $$
A typical value for $l$ is 26. Finally, define the histogram line as 
$$ H(t) = M(t) - S(t) $$
Figure \ref{fig:macd} illustrates features derived from the moving-average convergence divergence (MACD) metric. We compare a ``slow'' and ``fast'' exponential moving average, to capture where price levels have been recently relative to a slightly longer history.

\subsection*{A.3. Relative strength index}
Here we give one additional example of a momentum-based feature. The relative strength index (RSI) metric (\cite{Features}) compares the magnitudes of recent gains and losses, producing a number between 0 and 100. Define the average gains and losses at time $t$:
$$ AG(n,t) = \f{\sum_{i=1}^n \max\{0, R_{t-i} \}}{n} $$
$$ AL(n,t) = \f{\sum_{i=1}^n |\min\{0, R_{t-i} \}|}{n} $$
where we define the return at time $t$ as
$$ R_t = \f{P_t - P_{t-1}}{P_{t-1}} $$
Then define the first relative strength at time $t$ as
$$ FRS(n,t) = \f{AG(n,t)}{AL(n,t)} $$
and the smoothed relative strength at time $t$ as
$$ SRS(n,t) = \f{AG(n,t-1) \cdot (n-1) + \max\{0, R_t\} }{AL(n,t-1) \cdot (n-1) + |\min\{0, R_t\}| } $$
Now, the first relative strength index is defined as
$$ RSI_F = 100 - \f{100}{1 + FRS} $$
and the smoothed relative strength index is defined as
$$ RSI_S = 100 - \f{100}{1 + SRS} $$
A typical value for $n$ is 14 days.

\section*{Appendix B. SVM algorithm}
\subsection*{B.1. Problem setup}
Suppose data $\mathbf{x}_1,...,\mathbf{x}_n$ belong to two classes, with labels $t_1,...,t_n \in \{1,-1\}$ and feature transformation $\boldsymbol{\phi}$, and we model the predicted class assignment with
$$ y(\mathbf{x}) = \mathbf{w}^T \boldsymbol{\phi(\mathbf{x})} + w_0 $$
We seek parameters $\mathbf{w}, w_0$ such that
$$ t_i y(\mathbf{x}_i) \geq 0,\ \ \forall i=1,...,N$$
and that maximizes the minimal signed distance from any point to the boundary, known as the margin:
$$ \hat{\mathbf{w}}, \hat{w}_0 = \underset{\mathbf{w}, w_0}{\arg \max} \bigg( \frac{1}{\|\mathbf{w}\|} \underset{i \in \{1,...,N\}}{\min} \Big( t_i y(\mathbf{x}_i) \Big) \bigg) $$
Equivalently, we solve
$$ \hat{\mathbf{w}}, \hat{w}_0 = \underset{\mathbf{w}, w_0}{\arg \min} \frac{1}{2}\|\mathbf{w}\|^2  $$
such that
$$ t_i(\mathbf{w}^T\boldsymbol{\phi}(\mathbf{x}_i) + w_0) \geq 1,\ \forall i=1,...,n $$

\subsection*{B.2. Soft margin SVM}
Often, the dataset is not exactly linearly separable in the given feature space, and furthermore, a solution that exactly separates the classes may overfit to training noise. We allow some points to remain ``misclassified'' by defining a slack variable $\xi_i$ for each data point $\mathbf{x}_i$:
$$ \xi_i = \begin{cases} 
0, & t_i y(\mathbf{x}_i) \geq 1 \\
|t_i - y(\mathbf{x}_i)|, & t_i y(\mathbf{x}_i) < 1
\end{cases} $$
The slack variable represents the penalty incurred for violating the KKT conditions of the hard margin problem. If a point is exactly on the decision boundary, then $y(\mathbf{x}_i) = 0$ and $\xi_i = 1$. Points for which $\xi_i > 1$ are misclassified, and points for which $0 < \xi_i < 1$ are correctly classified but fall within the minimum margin. The soft margin problem can then be formulated as follows:
$$ \hat{\mathbf{w}}, \hat{w}_0 = \underset{\mathbf{w}, w_0}{\arg \min} \frac{1}{2}\|\mathbf{w}\|^2 + C \sum_{i=1}^n \xi_i $$
such that
$$ t_i(\mathbf{w}^T\boldsymbol{\phi}(\mathbf{x}_i) + w_0) \geq 1-\xi_i,\ \ \forall i=1,...,n $$
where $C > 0$ is a tuning parameter that controls the tradeoff between minimizing training error and tolerating misclassifications. A larger $C$ more heavily discourages violating the hard margin KKT conditions. In the limit $C \to \infty$, we recover the hard margin problem.
\\\\
We can then write the Lagrangian, with Lagrange multipliers $a_i,b_i \geq 0,\ \ \forall i=1,...,n$:
$$ \mathcal{L}(\mathbf{w},w_0,\mathbf{a}, \mathbf{b}) = \frac{1}{2}\|\mathbf{w}\|^2 + C \sum_{i=1}^n \xi_i - \sum_{i=1}^n a_i (t_i(\mathbf{w}^T\boldsymbol{\phi}(\mathbf{x}_i) + w_0) - 1 + \xi_i) - \sum_{i=1}^n b_i \xi_i $$
The stationarity conditions are:
$$ \frac{\partial \mathcal{L}(\mathbf{w},w_0,\mathbf{a},\mathbf{b})}{\partial \mathbf{w}} = 0 \implies \mathbf{w} = \sum_{i=1}^n a_i t_i \boldsymbol{\phi}(\mathbf{x}_i) $$
$$ \frac{\partial \mathcal{L}(\mathbf{w},w_0,\mathbf{a},\mathbf{b})}{\partial w_0} = 0 \implies 0 = \sum_{i=1}^n a_i t_i $$
$$ \frac{\partial \mathcal{L}(\mathbf{w},w_0,\mathbf{a},\mathbf{b})}{\partial \xi_i} = 0 \implies a_i = C - b_i,\ \forall i $$
Observe that $a_i = C - b_i,\ b_i \geq 0 \implies a_i \leq C$. Substituting back into the Lagrangian, we obtain the dual representation:
$$ \mathcal{D}(\mathbf{a}) = \sum_{i=1}^n a_i - \frac{1}{2} \sum_{i=1}^n \sum_{j=1}^n a_i a_j t_i t_j k(\mathbf{x}_i, \mathbf{x}_j) $$
where $k(\mathbf{x}_i, \mathbf{x}_j) = \boldsymbol{\phi}(\mathbf{x}_i)^T \boldsymbol{\phi}(\mathbf{x}_j) $. Notice that this expression is identical to the hard margin case. The difference between the soft and hard margin problems lies entirely in the constraints. For the soft margin problem, we have:
$$ 0 \leq a_i \leq C,\ \forall i=1,...,n $$
$$ \sum_{i=1}^n a_i t_i = 0 $$
The soft margin SVM must satisfy the following KKT conditions $\forall i=1,...,n$:
$$ a_i \geq 0 $$
$$ t_i y(\mathbf{x}_i) - 1 + \xi_i \geq 0 $$
$$ a_i [ t_i  y(\mathbf{x}_i) - 1 + \xi_i ] = 0  $$
$$ b_i \geq 0 $$
$$ \xi_i \geq 0 $$
$$ b_i \xi_i = 0 $$
Every data point satisfies either $a_i = 0$ or $t_i  y(\mathbf{x}_i) = 1 - \xi_i$. The points for which $a_i = 0$ are non-support vectors that do not contribute to the SVM solution at all. Only a sparse subset of points have $a_i \neq 0$: these are the support vectors that determine the solution.

\subsection*{B.3. Extension to multiple classes}
SVM easily generalizes to multiple classes. With $K$ classes, we run $K$ different one-against-all SVMs, where in each case the label is +1 if it belongs to a class and -1 if it doesn't. With functions $\hat{y}_1,...\hat{y}_K$, the final class assignment would then be $ \D \hat{k} \equiv \underset{k \leq K}{\max}\  \hat{y}_k(\mathbf{x}) $. One issue with this approach is we cannot guarantee that $\hat{y}_1,...\hat{y}_K$ are comparable, since they are trained on different tasks. Also, the datasets for each classifier are unbalanced, with the proportions of labels are roughly $1/K$ for +1 and $(K-1)/K$ for -1. Despite these shortcomings, one-against-all remains a popular method. Another approach would be to train $K(K-1)/2$ pairwise one-against-one classifiers, which is more computationally expensive, and does not result in better performance on our datasets.

\section*{Appendix C. Extended classification accuracy results}
\begin{figure}[h!]
	\centering
	\subfloat{
		\includegraphics[scale=0.8]{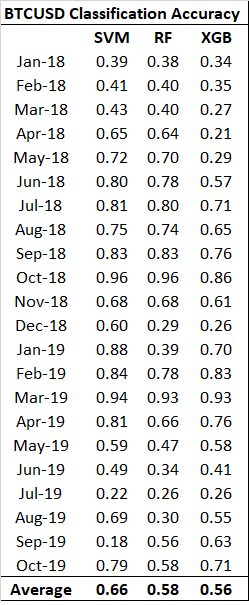}
	}\ \ \ 
	\subfloat{
		\includegraphics[scale=0.8]{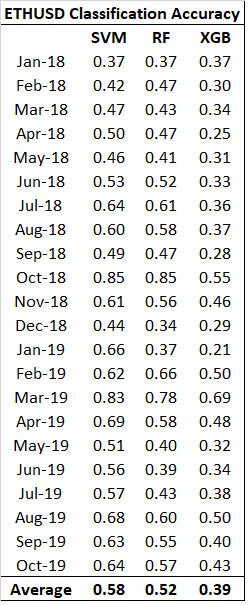}
	}\ \ \ 
	\subfloat{
		\includegraphics[scale=0.8]{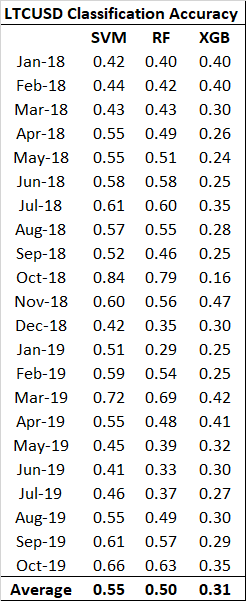}
	}
	\caption{\textit{These tables compare overall the classification accuracy of SVM, random forest, and XGBoost for the 3 crypto assets we tested. SVM consistently performs better than the other 2 classifiers, and also performs significantly better than chance. It is worth noting that some of these figures may overstate the effectiveness of our models, since the classes are highly imbalanced, as described in Section \ref{problem}. Thus, when certain months have very high overall accuracy, this may reflect the fact that the vast majority of points in that month fall into class $c_3$, i.e. we rarely make any trades.}}
	\vspace*{4in}
	\label{fig:overall_accuracy}
\end{figure}

Figure \ref{fig:overall_accuracy} shows the average monthly overall classification accuracy of SVM, random forest, and XGBoost from Jan 2018 to present. Because our model only trades a fraction of the time, these metrics are less relevant than PPV and NPV, as we described in Section \ref{accuracy}.

\section*{Appendix D. Volatility vs. Returns}
\begin{sidewaysfigure}[]
	\centering
	\includegraphics[scale=0.8]{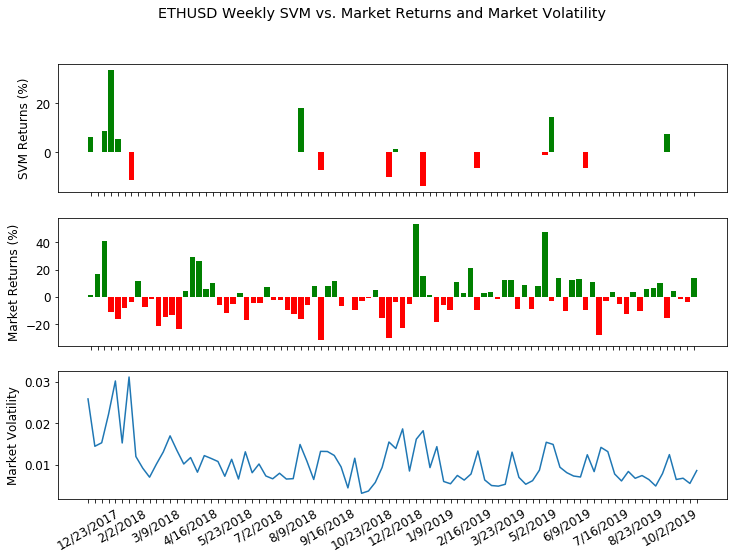}
	\caption{\textit{This figure compares the weekly returns for SVM and the market in ETHUSD, along with rolling weekly market volatility. We see that our strategy is more likely to trade when market volatility is high, and often does not trade at all in low volatility periods, when it is more ``uncertain'' of any price moves. Our strategy is also relatively uncorrelated with the market.}}
	\label{fig:ETH_weekly_returns}
\end{sidewaysfigure}

Figure \ref{fig:ETH_weekly_returns} compares weekly SVM and market returns for ETHUSD, along with rolling weekly market volatility. As we described in Section \ref{bitcoin}, we suspect that our strategy trades more in periods of higher volatility, which is more clearly demonstrated at a weekly granularity.

\end{document}